\documentclass[twocolumn]{aastex631}

\shorttitle{The Missing Link: Testing GCE Models}
\shortauthors{Coria et al.}
\graphicspath{{./}{figures/}}

\begin{document}

\title{The Missing Link: Testing Galactic Chemical Evolution Models with the First Multi-Isotopic Abundances in Solar Twin Stars}




\author[0000-0002-1221-5346]{David R. Coria}
\affiliation{Department of Physics $\&$ Astronomy, University of Kansas, Lawrence, KS, USA}

\author{Ian J. M. Crossfield}
\affiliation{Department of Physics $\&$ Astronomy, University of Kansas, Lawrence, KS, USA}

\author{Joshua Lothringer}
\affiliation{Department of Physics $\&$ Astronomy, Johns Hopkins University, Baltimore, MD, USA}

\author{Becky Flores}
\affiliation{Department of Physics $\&$ Astronomy, Georgia State University, Atlanta, GA, USA}

\author{Nikos Prantzos}
\affiliation{Institut d'Astrophysique de Paris, Paris, FR}

\author{Richard Freedman}
\affiliation{SETI Institute, Mountain View, CA, USA}

\begin{abstract}
We present the first isotopic abundances of both $^{13}$CO and C$^{18}$O in solar twin stars and test the results against several galactic chemical evolution (GCE) models with different nucleosynthesis prescriptions. First, we compare M-band spectra from IRTF/iSHELL to synthetic spectra generated from custom solar atmosphere models using the PHOENIX atmosphere code. Next, we compare our calculated abundances to GCE models that consider isotopic yields from massive stars, asymptotic giant branch (AGB) stars and fast-rotating stars. The $^{12}$C/$^{13}$C ratios determined for this sample of solar twins are consistent with predictions from the selected GCE models; however, the $^{16}$O/$^{18}$O ratios tentatively contradict these predictions. This project constitutes the first in a stellar chemical abundance series seeking to: (1) support the James Webb Space Telescope (JWST) as it characterizes exoplanet atmospheres, interiors, and biosignatures by providing host star abundances (2) identify how unexplored stellar abundances reveal the process of galactic chemical evolution and correlate with star formation, interior, age, metallicity, and activity; and (3) provide improved stellar ages using stellar abundance measurements. By measuring elemental and isotopic abundances in a variety of stars, we not only supply refined host star parameters, but also provide the necessary foundations for complementary exoplanet characterization studies and ultimately contribute to the exploration of galactic, stellar, and planetary origins and evolution. 
\end{abstract}



\section{Introduction} \label{sec:intro}

In stellar and planetary astrophysics, one of the most important areas of study is the chemical composition of the object in question. Until recently, large-scale photometric surveys could determine only the most fundamental stellar parameters such as mass, radius, luminosity, and temperature. While these parameters are important, they do not provide the context necessary for examining the chemical evolution of our galaxy, nor do they provide a baseline for exploring planet formation mechanisms, exoplanet atmospheres, and exoplanet interiors. For this, we need spectroscopy and stellar chemical abundances. Since generations of stars produce different elements, elemental and isotopic abundance ratios provide information on the age of a system and can be used as a sort of “cosmic clock”. These measurements are used to perform key tests of mixing mechanisms inside stars and serve as powerful diagnostic tools to constrain chemical enrichment, galaxy formation and evolution models when used alongside accurate stellar ages, distances, and kinematics \citep{Jackson:2021, Romano:2017}. 

On the larger, galactic scale, abundance surveys have revealed interesting trends between particular stellar elemental abundances, stellar age, and various stellar populations throughout the galaxy \citep{brewer:2016, adibekyan:2018, botelho:2020, nissen:2015, nissen:2020, delgado_mena:2019, delgado_mena:2021}. Other studies use stellar abundances to reconstruct the chemical history of the Milky Way \citep{jofre:2017, Jackson:2021} and to track down our Sun's long-lost siblings \citep{adibekyan:2018}. Some stellar abundances can even be used to distinguish exoplanet host stars from the general galactic stellar population \citep{brewer:2016a, swastik:2022, delgado_mena:2021, teske:2019}. Although galactic chemical evolution (GCE) models have already successfully reproduced present-day cosmic abundances of many elements (e.g. \cite{kobayashi:2011, prantzos:2018, Romano:2017}), new stellar abundance measurements allow for modelers to test the adopted stellar nucleosynthetic yields for which there is little observational data. Doing so will reveal whether the current understanding of stellar yields and chemical evolution models is truly accurate or simply appears accurate by coincidence. The $^{12}$C/$^{13}$C ratio is most commonly observed in giant stars where it is well known that various phenomena alter the initial $^{12}$C/$^{13}$C ratio. However, such comparisons also depend on stellar evolution and interior models. To complement these data, GCE model predictions should also be compared to $^{12}$C/$^{13}$C ratios measured in unevolved stars, like FGKM dwarf stars, that have preserved the initial isotope ratio in their envelopes. All of the GCE model studies mentioned in this paper \citep{kobayashi:2011, Romano:2017, prantzos:2018} issue a call for dwarf star isotopic abundance ratios, particularly for those isotope ratios (e.g. CNO \citep{Romano_2022}, Mg, Si, Ti) that we cannot trust from giant stars. Since these publications, the $^{12}$C/$^{13}$C ratio has been measured in solar twin stars \citep{botelho:2020}, M-dwarf stars \citep{crossfield:2019a}, and even in sub-stellar objects like brown dwarfs and directly imaged exoplanets \citep{zhang:2021b, zhang_2021a}.

On a smaller, planetary-system scale, stellar elemental abundances provide a glimpse into planetary formation and migration mechanisms as well as planetary chemical composition. Since stellar atmospheres evolve slowly, the elemental abundances of exoplanet hosts tend to reflect the composition of their planet-forming disks \citep{brewer:2016} and have the potential to yield constraints on planet formation processes and, in turn, even the physical properties of exoplanets themselves \citep{bedell:2018}. These chemical signatures, observed in the photospheres of stars, can be traced back to a planet's sequestration of heavy elements during planet formation or later in the system's history when a host star accretes planetary material. Thus, the presence, absence, and composition of planets could ideally be inferred from minute differences in the abundances between stars \citep{gaidos:2015}. 

Recent chemical abundance surveys explore the implications for planet formation of refractory element abundance (including C, O, Mg, and Si) and overall metallicity for thousands of solar analog stars and others within the local solar neighborhood \citep{fortney:2012, brewer:2016, bedell:2018, teske:2019, Nibauer:2021, swastik:2021}. Studying the chemical composition of carbonaceous chondrites within our own solar system \citep{anders:1989, braukmuller:2018} and the photospheres of nearby polluted white dwarfs \citep{harrison:2018, bonsor:2021, putirka:2021} provide another means of indirectly inferring the composition of planetary material and reinforce the link between stellar abundances and exoplanet composition. 

Other studies explore the possibility of using elemental abundance ratios like the carbon-to-oxygen ratio \citep{Oberg_2011, reggiani:2022, seligman:2022}, refractory-to-volatile ratios \citep{Hands_2021, Lothringer_2021, Welbanks_2019}, or isotopic abundance ratios like $^{12}$C/$^{13}$C \citep{zhang:2021b, zhang_2021a} to constrain a planet's formation location relative to stellar ``snowlines". As we enter the era of the space-based James Webb Space Telescope and ground-based ELTs, we also prepare to observe exoplanet atmospheres in unprecedented detail: from massive, accreting super-Jupiters like TYC 8998-760-1 b down to super-Earths and smaller terrestrial planets like those in the TRAPPIST-1 system. Contemporary and future researchers use these planetary and stellar abundances to model exoplanet atmospheres, to infer the structure and composition of the exoplanet's interior, and to understand how atmospheres and interiors co-evolve over time \citep{Madhusudhan:2012, Unterborn:2014, brewer:2016, unterborn:2017, Lincowski:2018, lincowski:2019}. 

Clearly, there is plenty of work being done to derive stellar elemental abundances and explore their relationship to planet formation and galactic chemical evolution, however, the same cannot be said for isotopic abundances. The present-day isotopic abundance database contains only a handful of measurements from giant stars and even fewer from dwarf stars and is currently barring progress to test isotopic abundances against GCE models and planet formation mechanisms.

In this paper, we present the first $^{12}$C/$^{13}$C and $^{16}$O/$^{18}$O ratio measurements made in solar twin stars using infrared fundamental CO band features. By developing a body of precise stellar isotopic abundance profiles, we aim to identify the “missing link” between GCE model predictions and observations, and to reveal the physical phenomena responsible for present-day abundances in the process. 

In Section \ref{sec:background}, we provide an overview of CO isotopic abundances including: isotope production in AGB stars, archival isotope measurements in giant stars, pilot studies in dwarf stars and the challenges faced, the potential of isotopic abundances to act as stellar chronometers, and also their role in constraining GCE models. In Section \ref{sec:sample_obs}, we present our solar twin sample (including fundamental stellar parameters and elemental abundances), describe our observations using IRTF/iSHELL \cite{rayner:2016} and spectral reduction. Section \ref{sec:analysis} contains a description of our PHOENIX stellar models and our isotopic abundance analysis methodology. In Section \ref{sec:results}, we present our $^{12}$C/$^{13}$C and $^{16}$O/$^{18}$O ratio measurements for our solar twin sample and compare them to GCE models \citep{kobayashi:2011, Romano:2017, prantzos:2018} and archival $^{12}$C/$^{13}$C measurements \citep{botelho:2020}. Finally, in Section \ref{sec:conclusions} we provide a brief commentary of our infrared, CO-based isotopic abundance analysis and present an overview of future stellar targets and other isotopic abundance ratios of interest.

\section{Background \& Motivation} \label{sec:background}
\subsection{Carbon and Oxygen Isotope Production Mechanisms}
As low-to-intermediate mass stars (1 $\leq M_\odot \leq$ 8) enter the late stages of their evolution, they play an important role in the chemical enrichment of the ISM, particularly in regard to minor isotope production. Asymptotic Giant Branch stars, or AGB stars, constitute the late phase in the evolution of these low-to-intermediate mass stars. Depending on time and location within the galaxy, up to 50\% of the material returned to the ISM by dying stars comes from AGB stars, \citep{goswami:2014} and the many grain species formed in their cool circumstellar envelopes contribute greatly to the dust population of the ISM \citep{gehrz:1989}. Therefore, understanding the nucleosynthetic processes and yields from these stars is crucial to chemical evolution modeling efforts. AGB star isotopic abundance ratios are great probes of internal mixing processes and their effect on stellar nucleosynthesis as different dredge-up events mix the burning material within the stellar envelope. During their red giant branch (RGB) phase, stars undergo a convective mixing process called the first dredge-up (FDU). This process carries nuclei from internal layers to the stellar surface. Because these internal layers are affected by partial CNO cycling which produces minor CNO isotopes like $^{13}$C, this leads to a decrease in the atmospheric $^{12}$C/$^{13}$C ratio with respect to main sequence values of $\sim90$ down to $\sim20-30$. The FDU leaves $^{16}$O abundances unaltered, increases the $^{17}$O abundance by about 50\%, and slightly reduces $^{18}$O \citep{abia:2017}. A second dredge-up event occurs during the early-AGB phase of the more massive stars (M $\geq 4 M_\odot$) where $^4$He and $^{14}$N are brought to the surface, but this does not modify the CNO isotope ratios significantly. The third dredge-up (TDU) event that occurs during the star's main AGB phase, however, mixes products of He burning into the envelope. Since $^{12}$C is the main product of He burning, the stellar surface becomes enriched in $^{12}$C and the $^{12}$C/$^{13}$C ratio increases past the previous FDU value \citep{abia:2017}. The amount of carbon in the envelope may eventually exceed the amount of oxygen, and so the AGB star becomes an AGB carbon star.

$^{12}$C, the main carbon isotope, is a primary product of the triple-$\alpha$ process that occurs during a star’s helium-burning phase and drives mixing between the processed core and outer envelope in low to intermediate-mass AGB stars. $^{12}$C from the hydrogen-burning shell in intermediate and massive stars is responsible for producing all of the nuclei involved in the CNO cycle: $^{13}$C, $^{15}$N, and $^{17}$O. While $^{12}$C is significantly produced by low mass stars (1-4 M$_{\odot}$), $^{13}$C is produced mainly in intermediate mass stars and massive stars via three main production mechanisms: (1) the CNO cycle resulting from H-burning; (2) $^{12}$C burning in low metallicity, fast-rotating massive stars; and (3) proton-capture nucleosynthesis in AGB stars \citep{botelho:2020}. State-of-the-art GCE models investigate the connection between stellar rotation and internal mixing mechanisms by implementing nucleosynthetic yields from massive, fast-rotating stars \citep{prantzos:2018, romano:2019}. 

Synthesis of the primary oxygen isotope, $^{16}$O, is well understood since it is a primary product of stellar evolution. It is produced exclusively by massive stars from $^{12}$C via $\alpha$-capture at the end of a star’s helium burning phase. Lower mass stars eject only the initial $^{16}$O of their envelope and thus do not contribute to the $^{16}$O enrichment of the galaxy. In fact, both the $^{16}$O and $^{18}$O stellar yields are often negative in intermediate-mass asymptotic giant branch (AGB) stars because they are destroyed by proton-capture nucleosynthesis during CN burning \citep{kobayashi:2011}. Production of the secondary oxygen isotopes, $^{17}$O and $^{18}$O, depends on pre-existing seed nuclei. The lighter of the two, $^{17}$O, is mainly produced through CNO burning of hydrogen into helium via the reaction: $^{17}$F $\rightarrow$ $^{17}$O + e$^{+}$ + neutrino, followed by $^{17}$O + $^{1}$H $\rightarrow$ $^{14}$N + $^{4}$He. The latter reaction leads to the production of the next heavier oxygen isotope. The heavy oxygen isotope, $^{18}$O, is primarily produced from $\alpha$-captures on $^{14}$N, left over from CNO cycling, which occurs during the initial stages of helium burning \citep{meyer:2008, Nittler:2012}. $^{18}$O is also secondary product of the CNO cycle and explosive nucleosynthesis (T $\sim$ 10$^9$ K) in massive stars. It is also worth noting that more massive stars contribute significantly to the production of $^{16}$O and $^{18}$O since these isotopes require helium burning for production whereas low-mass stars contribute more to $^{17}$O synthesis which only requires hydrogen burning \citep{meyer:2008, Nittler:2012}. The evolution of the minor oxygen isotope abundances is not well understood.  Specifically, the balance between $^{18}$O depletion via CNO cycling and $^{18}$O production during helium burning is not well studied. 

\subsection{Isotopic CNO Abundances}

\subsubsection{Giant Star CNO Isotopic Abundances}
CNO isotopic abundances have previously been measured in several red giant stars and AGB carbon stars. Observations show that $^{16}$O/$^{17}$O/$^{18}$O ratios are in agreement with evolutionary model predictions \citep{Tsuji:2006, abia:2017} which confirms that oxygen ratios are well established using post-FDU values. Although the observed $^{12}$C/$^{13}$C ratios in these stars appear fairly diversified in the range of $\sim$5-50 with a majority in the $\sim$10-20 range, most appear lower than theoretical predictions and much closer to the CNO cycle equilibrium $^{12}$C/$^{13}$C ratio of $\sim3.5$ \citep{Tsuji:2006, Takeda:2019}. These low $^{12}$C/$^{13}$C ratios could, in theory, be explained by extra mixing processes during the RGB and AGB evolution phases, however this would also imply extreme oxygen and nitrogen isotope ratios of $^{16}$O/$^{17}$O $\geq$ 1000, $^{16}$O/$^{18}$O $\geq$ 2000, and $^{14}$N/$^{14}$N $\geq$ 104 \citep{Tsuji:2006, abia:2017}. Puzzling observations of AGB carbon stars with very low $^{12}$C/$^{13}$C ratios do not show these high oxygen and nitrogen ratios. Moreover, the stars that \textit{do} demonstrate extreme oxygen and nitrogen ratios have normal $^{12}$C/$^{13}$C ratios \citep{abia:2017}.

Carbon and oxygen isotope ratio measurements in giant stars are not as helpful to GCE models as dwarf star measurements. This is because giant-star nucleosynthesis significantly alters the initial ratio throughout the star’s evolution whereas dwarf stars tend to preserve their initial isotopic ratios in their atmospheres \citep{meyer:2008, Nittler:2012, prantzos:2018}. Thus, dwarf star abundances better reflect the chemical composition of their birth clusters and are overall a better gauge of chemical evolution over time. 

\subsubsection{Challenges to Isotopic Abundance Measurements in Dwarf Stars} \label{subsec:challenges}

Until recently, it has been notoriously difficult to detect isotopes in cool dwarf stars, hence the inadequate isotopic abundance catalog. Isotopic effects are prominent only in molecular, not atomic, lines; therefore, spectral molecular features are most prominently observed in stars such as M-dwarfs whose photospheres are cool enough to allow the formation of various molecular species \citep{tsuji:2016b}. However, molecular observations come with some challenges due to telluric absorption and the sheer density of absorption lines throughout the optical and infrared. This includes strong molecular absorption due to metal oxides and hydrides \citep{Koizumi:2020} such as TiO in the optical band or H$_2$O in the NIR band \citep{souto:2018} and the opacity due to millions of other molecular absorption lines that dominate an M-dwarf’s observed spectrum. These spectral features have hampered past efforts to identify desired isotopic lines necessary for a detailed chemical analysis and accurate atmospheric models \citep{veyette:2018}. In addition, determining isotopic abundance ratios in stars requires very high-resolution and high signal-to-noise spectra \citep{adibekyan:2018, Romano:2017}. It is rather difficult to detect isotopologue features in dwarf star spectra, and so the database of isotopic carbon and oxygen abundances is rather small.

\subsubsection{Dwarf Star Carbon and Oxygen Isotopic Abundance Ratios}

Despite these challenges, a few isotopolgue detections have been made in dwarf stars ranging from F-type stars down to M dwarfs. A pilot study used medium-resolution (R $\sim 20,000$) near-infrared spectra to measure the $^{12}$C/$^{13}$C ratio in a sample of M dwarf stars \citep{tsuji:2016b}. The medium resolution observations could not resolve the faint $^{13}$C$^{16}$O from stronger $^{12}$C$^{16}$O, but evidence for $^{13}$C$^{16}$O in these M dwarf spectra was reported for the first time. This study demonstrates the need for very high-resolution spectra, otherwise CO isotopologue lines become too blended to measure precise abundances. 

With this caveat in mind, later studies successfully measured the $^{12}$C/$^{13}$C in solar twins using $^{12}$CH and $^{13}$CH features in the optical band \citep{adibekyan:2018, botelho:2020}. These efforts relied on high-resolution (R $\sim$120,000) and high S/N optical HARPS spectra to make their measurements. This high resolution allowed them to discern weaker $^{13}$CH lines from the predominant $^{12}$CH lines. The $^{12}$C/$^{13}$C is used to look for stars from the same birth cluster as the Sun \citep{adibekyan:2018} and also to identify $^{12}$C/$^{13}$C trends with stellar isochrone age \citep{botelho:2020}. 

For cooler stellar and sub-stellar objects, the preferred method is to target the $^{12}$CO/$^{13}$CO ratio instead since this carbon isotope ratio is more readily detectable and attained from the ground using near- and mid-infrared observations \citep{molliere:2019}. Recent observations show that infrared observations of the CO fundamental rovibrational band at high resolution allow the observation of rarer isotopologues, even in dwarf stars, by resolving individual lines in the spectrum and providing sensitivity to much lower abundances of $^{13}$CO and C$^{18}$O \citep{crossfield:2019a}. This approach was used to measure the first multiple isotopic abundances of M-dwarfs in the mid-infrared at high resolution (R $\sim 60,000$): not merely the common $^{12}$C$^{16}$O but also the rarer species $^{13}$C$^{12}$O and $^{12}$C$^{18}$O \citep{crossfield:2019a}. Since CO is an oxygen-bearing molecule, unlike CH used in earlier studies, these CO isotopologue features allow us to measure the $^{16}$O/$^{18}$O in addition to the $^{12}$C/$^{13}$C ratio. This result shows that isotopic abundance analysis is not only possible via high resolution spectroscopy, but it is also the next logical step for many stellar targets.

In the following sections, we go into further detail describing our present understanding of elemental and isotopic abundances and how they may provide the ``missing link" between stellar photospheres and galactic chemical evolution.

\subsection{Chemical Abundances as Stellar Chronometers}
Investigations into the mechanisms responsible for chemical enrichment of the galaxy have led to the study of elemental abundances as ``stellar chronometers" that have the potential to be used to determine stellar ages. It is assumed that the chemical composition of low-mass dwarf stars remains unchanged throughout stellar lifetimes , and instead chemical abundances evolve in time through inheritance between stellar generations. Because older stellar populations synthesize metals and release them back into the ISM to be recycled into new stars, elemental abundances increase over time \citep{da_Silva:2012}. Past observational data has shown that there is no tight correlation between stellar ages and metallicities \citep{Holmberg:2007, Casagrande:2011}, therefore we have to delve beyond and explore the relation between individual elemental abundances and stellar age. Core-collapse (Type II) supernovae enriched the early galaxy with $\alpha$-process elements (C, O, Ne, Mg, Si, S, Ar, Ca, Ti) on a faster timescale than Type Ia supernovae could enrich the galaxy in iron-peak elements (e.g. Cr, Mn, Fe, Ni) \citep{kobayashi:2020}. This delay in iron-peak element enrichment produces chemical signatures like [$\alpha$/Fe] that serve as a good proxy for stellar age and a probe for the history of galactic chemical evolution \citep{Haywood:2013, delgado_mena:2019, kobayashi:2020}. In general, the [$\alpha$/Fe] abundance ratios and others formed mostly by Type II supernova increase with the age of a star \citep{nissen:2015, bedell:2018, delgado_mena:2019}. The ratios of s-process elements (Sr, Ba, Zr, Y, La), primarily produced in low-mass AGB stars, over Fe follow trends similar to the iron-peak elements and show a negative correlation with stellar age \citep{delgado_mena:2019}. The first major study of elemental abundances and their relation to age by \cite{da_Silva:2012} identified [Y/Mg], [Sr/Mg], [Y/Zn] and [Sr/Zn] as having significant trends as a function of age in a large sample of solar twins. Since this study, several other abundance ratios have been identified as potential chemical clocks and we discuss them in more detail below.

Among stellar chronometer candidates, the most widely discussed is perhaps the [Y/Mg] ratio. In the context of stellar nucleosynthesis, it makes sense that this abundance ratio is sensitive to stellar age. Massive stars produce and expel Mg into the ISM on a different timescale than intermediate-mass AGB stars produce and expel Y. [Y/Fe] notably decreases with stellar age contrary to other s-process elements with a rather steep negative slope (slope = -0.033 dex Gyr$^{-1}$) and [Mg/Fe] increases with stellar age (slope = +0.009 dex Gyr$^{-1}$) \citep{nissen:2015} and so [Y/Mg] acts as a sensitive probe of stellar age. Multiple studies have confirmed the trend of [Y/Mg] with age \citep{nissen:2015, Tucci_Maia:2016}, and their equations can determine a stellar age with $\sim$0.8 Gyr precision (for solar twin stars aged 0-10 Gyr) given a precise measurement of [Y/Mg] ($\pm 0.02$). The best stellar chronometer candidates are often ratios of light s-process elements (Sr, Ba, Zr, Y, La) over $\alpha$ elements (C, O, Ne, Mg, Si, S, Ar, Ca, Ti plus Zn and Al) \citep{delgado_mena:2019} because of the drastically different ``enrichment timescales". For more information on elemental abundance ratios that demonstrate a trend with stellar age in solar twin samples, refer to \cite{nissen:2015, Tucci_Maia:2016, Nissen:2016, spina:2018, delgado_mena:2019, Jofre:2020, nissen:2020}.

Since the enrichment timescales for major and minor isotope are analogous to those for the best stellar chronometer candidates, ratios of major to minor isotopes must also be considered. In fact, \cite{botelho:2020} have recently reported that the $^{12}$C/$^{13}$C ratio in solar twin stars seems to have a tentative correlation with metallicity but not quite with stellar age. It remains unknown how the $^{16}$O/$^{18}$O ratios behave with stellar age or metallicity solar twin stars because of the difficulty in detecting faint C$^{18}$O. In this paper, we finally explore how both the $^{12}$C/$^{13}$C and $^{16}$O/$^{18}$O ratio correlate with stellar age in a small sample of solar twin stars, but ultimately more dwarf star observations are necessary to test carbon and oxygen isotope ratios' performance as chemical clocks. 

\subsubsection{Solar Twin Galactic Chemical Evolution}
Solar twin stars also exhibit peculiar abundance signatures that divide the solar twin population into distinct age groups and help trace their origin within the galaxy. Examination of solar twin stars in the solar vicinity show that the ages of these stars are widely distributed from 0-10 Gyr \citep{tsujimoto:2021}. Stars of similar ages share elemental abundance patterns as anticipated by the chemical evolution individual elements over time \citep{bedell:2018}. The oldest group of solar twins (ages $\sim 8.7$ Gyr) has abundance patterns (C to Dy) quite similar to solar-metallicity galactic bulge stars. The super-solar abundances of this older group of solar twins is likely a result of the faster chemical enrichment that occurs closer to the galactic center. These abundance patterns further suggest that the oldest solar twins have migrated to the local solar neighborhood from birthplaces within the galactic bulge \citep{tsujimoto:2021}. The second group of solar twins with ages $\sim 5.9$ Gyr demonstrate abundance patterns nearly identical to the Sun, and this association of the Sun with slightly older solar twins reinforces the notion of a common birthplace much closer to the galactic center. The $^{12}$C/$^{13}$C and $^{16}$O/$^{18}$O ratios, if proven to have a significant trend with stellar age, would be powerful chemical clocks and stellar formation location tracers since they also exhibit prominent gradients with galactocentric radius \citep{romano:2019}.

Clearly there is plenty to explore in galactic chemical evolution and exoplanet systems using elemental and isotopic abundances. However, dwarf star isotopic abundance measurements are quite scarce despite being more reliable than giant star measurements. Currently, isotopic abundance analyses focus on solar twin stars of near-solar metallicity. It is crucial we expand the database to probe low-mass \textit{and} low-metallicity stars where GCE models lack observational constraints. Therefore, we present an isotopic carbon and oxygen abundance analysis of six bright, well-studied solar twins. This paper serves as a pilot study using infrared (M band) CO features to derive $^{12}$C/$^{13}$C and $^{16}$O/$^{18}$O ratios. Once applied to solar twin stars, the technique could be extended to K and M dwarf stars as well as low-metallicity stars to bridge the `observational' gap in GCE models.

\begin{deluxetable*}{ccccccccc}[h!p]
\tablenum{1}
\tablecaption{Solar Twin Parameters}
\tablewidth{0pt}
\tablehead{
\colhead{Stellar} & \colhead{Spectral} & \colhead{Brightness} & \colhead{Radius} & \colhead{Age} &\colhead{*T$_{eff}$} & \colhead{Stellar} & \colhead{*Model} & \colhead{*Model} \\
\colhead{ID} & \colhead{Type} & \colhead{K Mag} & \colhead{(R$_\sun$)} & \colhead{(Gyr)} & \colhead{(K)} & \colhead{log g} & \colhead{T$_{eff}$} & \colhead{log g}
}

\startdata
Sun & G2V & 5.08 & 1.0 & 4.567 $\pm\ $0.11 & 5780 & 4.44 & 5780 & 4.44\\
HIP 29432 & G4V & 5.301 & 0.95 & 5.51 $\pm\ $0.71 & 5758 $\pm\ $5 & 4.44 & 5780 & 4.44\\
HIP 42333 & G8V & 5.223 & 1.0 & 1.01 $\pm\ $0.52 & 5848 $\pm\ $8 & 4.50 & 5870 & 4.49\\
HIP 77052 & G2V & 4.300 & 0.96 & 3.67 $\pm\ $0.91 & 5683 $\pm\ $5 & 4.48 & 5690 & 4.39\\
HIP 79672 & G2V & 4.190 & 1.03 & 3.09 $\pm\ $0.40& 5814 $\pm\ $3 & 4.45 & 5780 & 4.44\\
HIP 85042 & G3V & 4.686 & 1.04 & 6.66 $\pm\ $0.62 & 5694 $\pm\ $5 & 4.41 & 5690 & 4.39\\
HIP 102040 & G5V & 4.921 & 0.96 & 2.42 $\pm\ $0.91 & 5838 $\pm\ $6 & 4.48 & 5870 & 4.49\\
\enddata
\tablecomments{Stellar Parameters. Stellar ID from the Hipparcos Catalogue (ESA 1997); Spectral type and K band magnitude from 2MASS \citep{cutri:2003}; Stellar radius from Gaia DR2 \citep{gaia_dr2_2018}; Stellar age, Stellar effective temperature *T$_{eff}$, and stellar surface gravity (log g) from \cite{dos_Santos:2016}; Solar K Mag from \cite{wilmer:2018}; Solar age from \cite{Bonanno_2002, Jacobsen_2008}. The final two columns contain the stellar effective temperature and surface gravity parameters adopted for each set of PHOENIX model spectra.}
\label{tab:param}

\end{deluxetable*}

\begin{deluxetable*}{cccccccc}[h!p]
\tablenum{2}
\tablecaption{Observational Parameters}
\tablewidth{0pt}
\tablehead{
\colhead{Parameter} & \colhead{HIP 29432} & \colhead{HIP 42333} & \colhead{HIP 77052} & \colhead{HIP 79672} & \colhead{HIP 85042} & \colhead{HIP 102040}
}

\startdata
Date [UT] & 2019-02-02 & 2019-03-29 & 2019-05-16 & 2019-05-16 & 2019-05-16 & 2019-05-16\\
Time [UT] & 07:48 - 08:23 & 06:14 - 06:52 & 07:44 - 08:01 & 08:25 - 08:53 & 10:53 - 11:30 & 14:31 - 15:03\\
iSHELL Mode & M1 & M1 & M1 & M1 & M1 & M1 \\
Slit & 0.375” × 15” & 0.375” × 15” & 0.375” × 15” & 0.375” × 15” & 0.375” × 15” & 0.375” × 15”\\
Integration Time [s] & 48.65 & 63.94 & 48.65 & 48.65 & 48.65 & 48.65\\
Co-adds & 3 & 3 & 3 & 3 & 3 & 3 \\
Exposures & 40 & 34 & 20 & 32 & 42 & 32 \\
Median S/N & 16.6 & 36.2 & 32.5 & 45.3 & 36.4 & 22.0 \\
\enddata
\tablecomments{Observational parameters for the six solar twins on IRTF/iSHELL \citep{rayner:2016}}
\label{tab:obs_param}
\end{deluxetable*}

\begin{figure*}[h!p]
    \centering
    \epsscale{0.95}
    \includegraphics[width=0.95\textwidth]{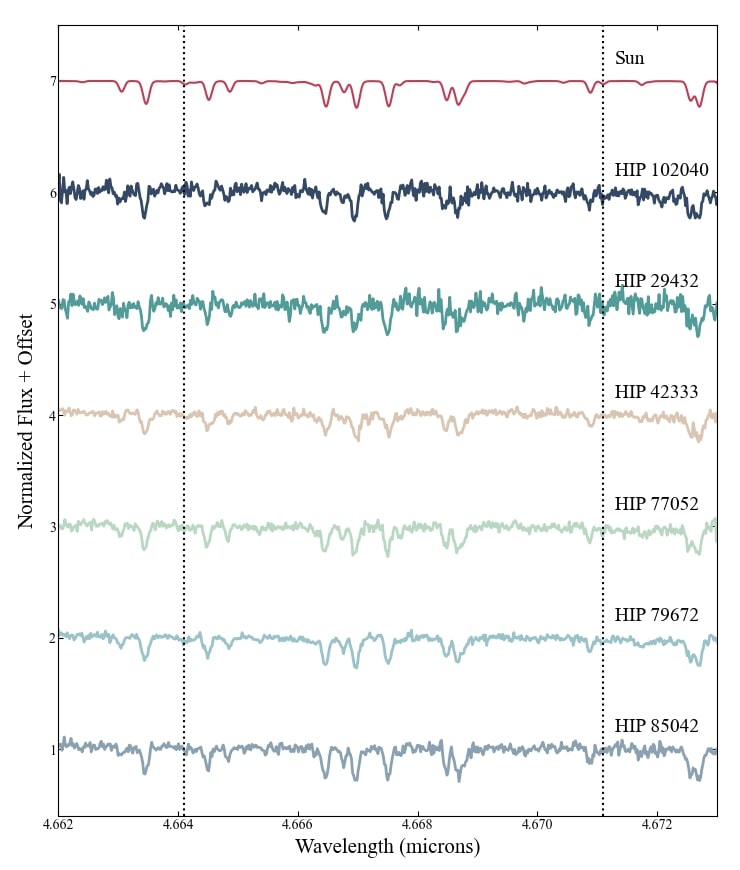}
    \caption{A small section of the normalized spectra for the Sun from the ACE-FTS Solar Atlas \citep{hase:2010} and our six solar twins from IRTF/iSHELL \citep{rayner:2016}. The spectra have been shifted to the same wavelength frame. The black dotted lines show the location of some $^{13}$CO features used in our abundance analysis.}
    \label{fig:Solar Twin Spectra}
\end{figure*}


\section{Sample and Observations} \label{sec:sample_obs}
\subsection{Solar Twin Sample}
We selected a sample of six bright solar twin stars with near-solar spectral types, radii, effective temperatures, metallicities, carbon abundances, and oxygen abundances (See Tables ~\ref{tab:param} and \ref{tab:age,ab,results}). The stellar ages used in this analysis are taken from \citep{dos_Santos:2016} and originally obtained by \cite{Tucci_Maia:2016} using Yonsei-Yale isochrones \citep{Yi:2001}. The solar twin ages range from 1-7 Gyr, but the metallicities are all within 0.15 dex from solar. The solar twin ages cover a large portion of the ``GCE vs. Time" space; however, as our sample is nearly solar in [Fe/H], the low-metallicity ([Fe/H $< -0.2$]) space remains unexamined by this sample. 

Three out of the six solar twins in this sample appear in multi-star systems: HIP 77052 is in a binary \citep{hirsch:2021}, HIP 85042 is in a triple system \citep{Riddle:2015}, and HIP 102040 has three optical companions \citep{Riddle:2015}. Out of the six solar twins, HIP 29432 is the only confirmed exoplanet host. HIP 29432 hosts a modestly-irradiated Neptune-mass planet with a semimajor axis of 0.55 AU \citep{fulton:2016}. 

\subsection{Observations}

We observed the six solar twins listed in Table ~\ref{tab:obs_param} during the 2019A semester using the iSHELL spectrograph \citep{rayner:2016} on the NASA Infrared Telescope Facility. Table \ref{tab:obs_param} lists the technical details of these observations. We obtained spectra at R = 70,000 (4.3 km/s) and mostly-continuous coverage from 4.52–5.24 $\mu$m. We reduced the raw iSHELL data using the SpeXTool Data Reduction package \citep{cushing:2004} which corrects for pixel-to-pixel variations and produces the wavelength calibration. We then compute spatial profiles and extract two one-dimensional spectra (one at each nod position). These two spectra are combined to produce a single spectrum for each star. 

We correct for telluric absorption in our spectra following the approach of \citep{crossfield:2019a}, using the following A0V standard stars (V Magnitude): HR 7891 (4.82), HR 2133 (6.075), HR 3314 (3.90), HR 5881 (3.53), HR 5881 (3.53), HR 6629 (3.75). Finally, we remove parts of the spectrum with obvious bad pixels and wherever S/N $<$ unity. In practice, the choice of S/N cut-off is not especially significant since low-S/N parts of the spectrum are appropriately de-weighted when we calculate our weighted-mean line profile for each isotopologue (see Section \ref{sec:analysis}).

\newpage
\section{Measuring Isotopic Abundances} \label{sec:analysis}
\subsection{Model Spectra}

To measure CO isotopologue abundances in our solar twin sample, we compare our observed spectra to synthetic spectra generated from custom solar atmosphere models derived from the PHOENIX atmosphere code \citep[Version 16;][]{husser:2013}\citep{Hauschildt_1999}. Our PHOENIX model atmospheres contain 64 vertical layers, spaced evenly in log-space on an optical depth grid from $\tau$ = 10$^{-8} - 1000$ spanning $1.0 - 10^6$ nm. In our observed wavelength range, the models were sampled at least every 0.01 nm. In contrast to the M-dwarf synthetic spectra from our similar analysis \cite{crossfield:2019a}, the solar-like models for this analysis were only run with H I, He I, and He II in NLTE to reduce computation time. For both the $^{13}$CO and the C$^{18}$O analysis, we generate synthetic spectra with $^{13}$C and $^{18}$O enrichments of 3$\times$, 1.78$\times$, 1$\times$, 0.56$\times$, 0.3$\times$, and 0$\times$ solar. We use a CO line list \citep{goorvitch:1994} that contains lines for $^{12}$C$^{16}$O, $^{13}$C$^{16}$O, $^{12}$C$^{18}$O, and other CO isotopologues. This older line list is in good agreement with newer line lists of the CO fundamental band. These models are run for effective temperatures (in K) of 5690, 5780, and 5870 and log$_g$ (cgs) of 4.39, 4.44, and 4.49 respectively. Each solar twin's isotopologue abundance analysis is performed with the set of model spectra that best fits that star's effective temperature and log g (refer to Table \ref{tab:param}).

\subsection{Line Selection} \label{subsec:line_selection}
We measured our isotopic abundances following the process described in \cite{crossfield:2019a}. In this study, we use relatively strong, isolated $^{13}$CO and C$^{18}$O lines to determine $^{13}$C and $^{18}$O isotopic abundances. Identifying these lines requires a focus on the highest S/N regions of the stellar spectra around 4.6–4.7 microns. Here, tellurics are relatively weak and the stellar spectra are dominated by $^{12}$C$^{16}$O, $^{13}$C$^{16}$O, and $^{12}$C$^{18}$O lines from the CO fundamental rovibrational band. We have compiled a list of the strongest $^{13}$CO and C$^{18}$O lines within our desired wavelength range using the HITRAN database \citep{gordon_hitran2016_2017}. The HITRAN line lists used for CO line identification are based on data from \cite{goorvitch:1994} line lists used in our PHOENIX models. In this wavelength range, there are 41 $^{12}$CO Lines, 34 $^{13}$CO Lines, and 32 C$^{18}$O lines, but not all of them are included in the abundance calculation. Some lines are obscured by strong telluric absorption lines, some fall beyond outside the wavelength coverage in our IRTF/iSHELL spectra, and some weaker isotopologue lines are overshadowed by stronger absorption from $^{12}$C$^{16}$O and other molecular species. Please refer to our machine-readable line lists and the associated README file for more details on our line selection process.

\begin{figure*}[h!tp]
    \centering
    \includegraphics[width=0.90\textwidth]{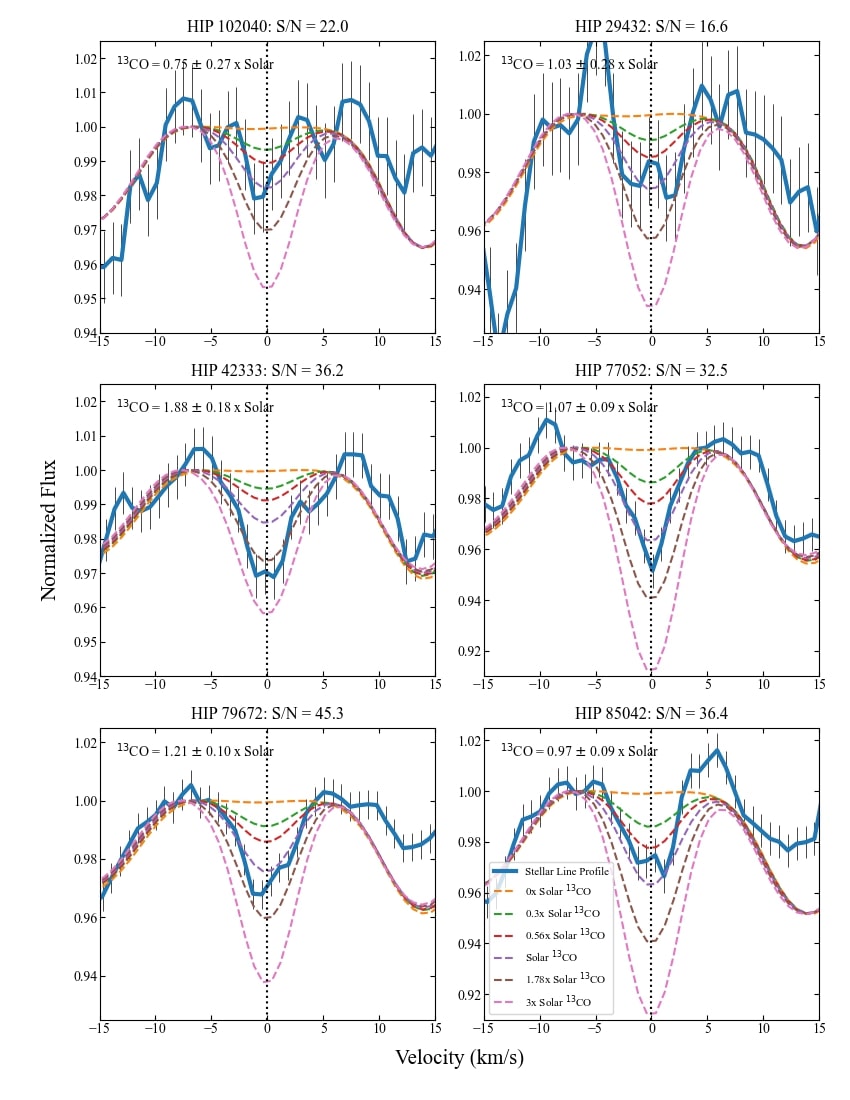}
    \caption{Stacked $^{13}$CO line profiles for our sample. The line profiles represent the weighted average of several M-band $^{13}$CO lines for the observed stellar spectrum (in blue) and multiple PHOENIX model spectra of varying $^{13}$CO abundances. We interpolate between these models using $\chi^2$ minimization to determine the stellar $^{13}$CO abundance relative to solar (See analysis example in Fig.~\ref{fig:Solar Analysis}; results shown in Table~\ref{tab:age,ab,results}).}
    \label{fig:13_CO Line Profiles}
\end{figure*}

\begin{figure*}[h!tp]
    \centering
    \includegraphics[width=0.90\textwidth]{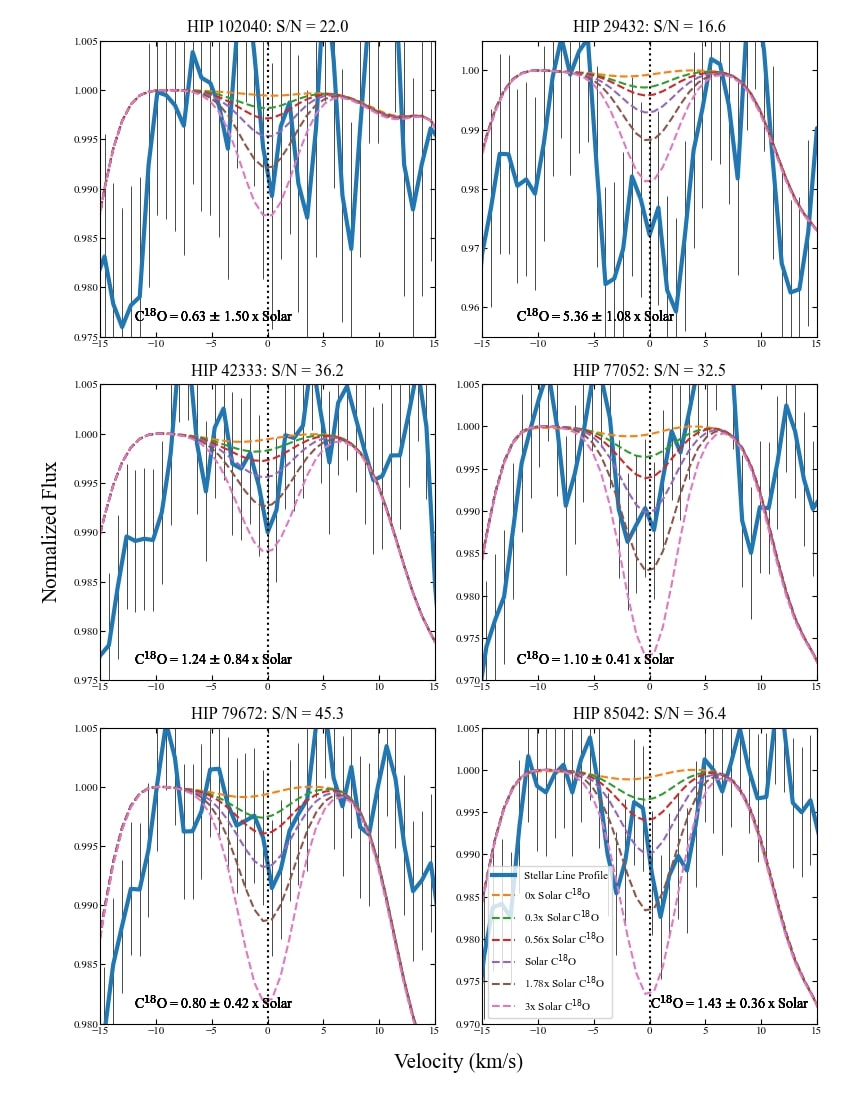}
    \caption{Stacked C$^{18}$O line profiles for our sample. The line profiles represent the weighted average of several M-band C$^{18}$O lines for the observed stellar spectrum (in blue) and multiple PHOENIX model spectra of varying C$^{18}$O abundances. We interpolate between these models using $\chi^2$ minimization to determine the stellar C$^{18}$O abundance relative to solar ($\chi^2$ minimization shown in Fig.~\ref{fig:Solar Analysis}; results shown in Table~\ref{tab:age,ab,results}).}
    \label{fig:C_18_O Line Profiles}
\end{figure*}

\subsection{Abundance Calculation} \label{subsec:ab_calc_Sun_ex}
Atomic abundances are often calculated using only a few absorption lines with relatively high S/N. In our case, however, it is important to note that most of these spectral lines belonging to the isotopologues of interest have low statistical significance and are barely visible by eye when considered individually. We thus create a single line profile of each CO isotopologue, for each solar twin in our sample by taking all useable absorption lines for each isotopologue and combining them into a single, high S/N line profile. We create the single line profile by taking the weighted mean, after continuum-normalizing, of each spectral line. This produces the stacked absorption profiles shown in Fig.~\ref{fig:13_CO Line Profiles} and Fig.~\ref{fig:C_18_O Line Profiles}. We then create corresponding line profiles for the synthetic stellar models detailed above, using the same set of lines. The isotopic abundances for each CO isotopologue are then determined by comparing the stacked absorption lines from the observed spectrum to the those of the synthetic models.

\begin{figure*}[hp]
    \centering
    \epsscale{1.0}
    \plottwo{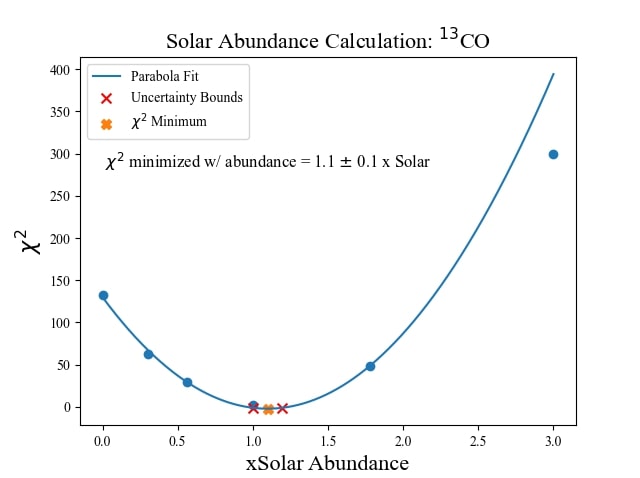}{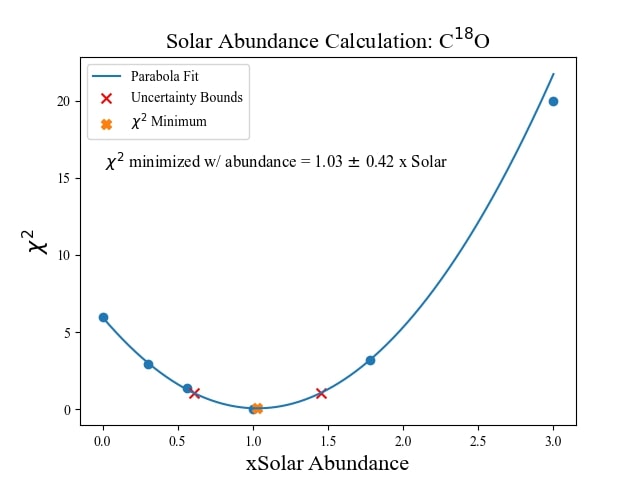}
    \caption{Example of our analysis approach for the infrared solar spectrum from the ACE-FTS Solar Atlas \citep{hase:2010}. Here we show the six $\chi^2$ values calculated between the solar line profile and each model of varying $^{13}$CO and C$^{18}$O abundances. We fit a parabola to the  $\chi^2$ values ($\leq 175$) and assign the minimum as the best-fit isotopologue abundance relative to solar. We measure $1.10 \pm 0.10$ and $1.03 \pm 0.42$ x Solar abundances for $^{13}$CO and C$^{18}$O respectively.}
    \label{fig:Solar Analysis}
\end{figure*}

We measure the $^{13}$C/H and $^{18}$O/H abundance ratios for each solar twin by a $\chi^2$ analysis, which represents how well the model line profile fits the observed solar twin line profile. We calculate $\chi^2$ over the velocity range $\Delta V$ ($6 < \Delta V < 12$) km/s centered at the line profile center V = 0. Because we want to minimize $\chi^2$ to identify the best fit abundance value, the set of $\chi^2$ values (only including values $\leq 175$) is fit with a parabola and the minimum is then calculated, giving the best-fit isotopologue abundance. We infer 1$\sigma$ confidence intervals using the region where $\Delta\chi^2 \leq 1$ \citep{avni:1976}. Fig.~\ref{fig:Solar Analysis} shows an example of this approach for the Sun (further described below). The final $^{13}$CO and C$^{18}$O abundances are shown in Table ~\ref{tab:age,ab,results}.

To test the accuracy of our $^{13}$C/H and $^{18}$O/H measurements within our solar twin sample and the consistency across different CO line lists, we also repeat this analysis using an infrared spectrum of the Sun from the ACE-FTS Solar Atlas \citep{hase:2010}. Ideally, if we measure the $^{13}$CO and C$^{18}$O abundances relative to solar \textit{in a spectrum of the Sun}, our analysis should find an abundance of 1.0 x Solar for each CO isotopologue. This is, nearly, what we observe. Using this solar spectrum with uncertainties adopted from our highest S/N spectrum (HIP 79672; see Table~\ref{tab:obs_param}), we measure $1.10 \pm 0.04$ and $1.03 \pm 0.42$ x Solar abundances for $^{13}$CO and C$^{18}$O respectively. The slight $^{13}$CO overestimate appears in the solar twin analysis even after the $\chi^2 \leq 175$ cutoff. These results indicate that our systematic and modeling errors are $\leq 10 \%$ for both isotopologues. We discuss this issue further in Section ~\ref{subsec:botelho_comp}.

\section{Results} \label{sec:results}
Our $^{13}$CO, C$^{18}$O, $^{12}$C/$^{13}$C, and $^{16}$O/$^{18}$O measurements are summarized in Table~\ref{tab:age,ab,results} for our six-solar-twin sample and the Sun. Similar to the analysis in \cite{botelho:2020}, we calculate a weighted least-squares linear fit for our $^{12}$C/$^{13}$C ratios to identify tentative trends. The fit demonstrates a decrease in the $^{12}$C/$^{13}$C ratio over metallicity (slope = -157 $\pm$ 82 dex$^{-1}$). Similarly, we see the $^{12}$C/$^{13}$C ratio decrease over time with a slope of -6.65 $\pm$ 2.15 Gyr$^{-1}$. A weighted linear fit of our oxygen ratios demonstrates an increasing trend over both metallicity and time: slope = +2438 $\pm$ 713 dex$^{-1}$ and +75.55 $\pm$ 56.8 Gyr$^{-1}$ respectively. Although the \cite{botelho:2020} data suggest a systematic increase in $^{12}$C/$^{13}$C with [Fe/H] but not with time, our sample does not seem to significantly favour either an increase or decrease. 

\begin{figure*}[h!tp]
    \centering
    \includegraphics[width=1.0\textwidth]{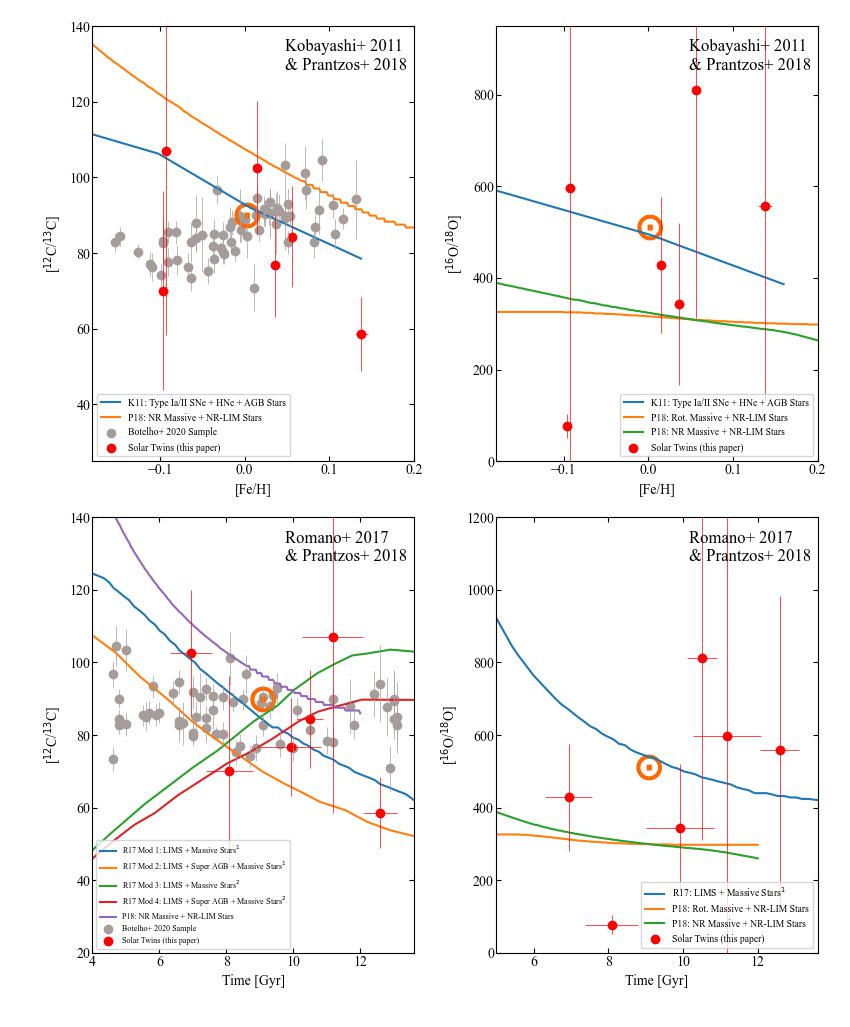}
    \caption{Comparison of our six solar twin $^{12}$C/$^{13}$C and $^{16}$O/$^{18}$O ratios to GCE models \citep{kobayashi:2011, Romano:2017, prantzos:2018} and archival solar twin measurements \citep{botelho:2020}.}
    \label{fig:GCE Model Comparison}
\end{figure*}

We will now discuss how our new carbon and oxygen isotope ratios compare to various GCE models. We will also examine how these solar twin carbon isotope ratio measurements compare to archival measurements. A visual comparison of our measurements to several GCE models is shown in Fig. \ref{fig:GCE Model Comparison}. The top two panels show $^{12}$C/$^{13}$C vs. [Fe/H] (left) and $^{16}$O/$^{18}$O vs. [Fe/H] (right). The bottom two panels show $^{12}$C/$^{13}$C vs. Time (left) and $^{16}$O/$^{18}$O vs. Time (right).

Figure \ref{fig:GCE Model Comparison} shows our solar twin $^{12}$C/$^{13}$C and $^{16}$O/$^{18}$O ratio measurements compared to GCE models and archival measurements ($^{13}$CO only). The top four panels show the predicted evolution of $^{12}$C/$^{13}$C (left) and $^{16}$O/$^{18}$O (right) ratios over stellar metallicity using GCE models from \cite{kobayashi:2011, prantzos:2018}. Similarly, the bottom two panels show the predicted evolution of $^{12}$C/$^{13}$C (left) and $^{16}$O/$^{18}$O (right) ratio over time using GCE models from \cite{Romano:2017}. The orange circle represents Solar values from \cite{ayres:2013} and the gray points represent the values of the \cite{botelho:2020} solar twin sample. The red points represent the solar twin abundances measured in this paper. Our measurements agree somewhat with GCE predictions for $^{12}$C/$^{13}$C, but are less obvious for $^{16}$O/$^{18}$O. We elaborate on these GCE models and the comparison below.

\begin{deluxetable*}{ccccccccc}[hp]
\tablenum{3}
\tablecaption{$^{12}$C/$^{13}$C and $^{16}$O/$^{18}$O Abundances}
\tablewidth{0pt}
\tablehead{
\colhead{} & \colhead{Age} & \colhead{[Fe/H]} & \colhead{[C/H]} & \colhead{[O/H]} & \colhead{$^{13}$CO/H} & \colhead{C$^{18}$O/H} & \colhead{Measured} & \colhead{Measured}\\
\colhead{ID} & \colhead{(Gyr)} & \colhead{(dex)} & \colhead{(dex)} & \colhead{(dex)} & \colhead{(x Solar)} & \colhead{(x Solar)} & \colhead{${^{12}}$C/${^{13}}$C} & \colhead{${^{16}}$O/${^{18}}$O}
}

\startdata
Sun & *4.603 & 0.0 & 0.0 & 0.0 & 1.10 $\pm$ 0.10 & 1.03 $\pm$ 0.42 & $*91.4 \pm 1.3$ & $*511 \pm 10$\\
HIP 29432 & 5.51 & -0.096 & -0.103 & -0.092 & 1.03 $\pm$ 0.28 & 5.36 $\pm$ 1.08 & $70 \pm 26$ & $77 \pm 26$\\
HIP 42333 & 1.01 & 0.138 & -0.047 & -0.130 & 1.88 $\pm$ 0.18 & 1.24 $\pm$ 0.84 & $59 \pm 10$ & $558 \pm 426$\\
HIP 77052 & 3.67 & 0.036 & 0.048 & 0.104 & 1.07 $\pm$ 0.09 & 1.10 $\pm$ 0.41 & $77 \pm 13$ & $344 \pm 177$\\
HIP 79672 & 3.09 & 0.056 & 0.037 & 0.078 & 1.25 $\pm$ 0.10 & 0.80 $\pm$ 0.42 & $84 \pm 13$ & $811 \pm 499$\\
HIP 85042 & 6.66 & 0.015 & 0.081 & 0.131 & 0.97 $\pm$ 0.09 & 1.43 $\pm$ 0.36 & $103 \pm 17$ & $428 \pm 148$\\
HIP 102040 & 2.42 & -0.093 & -0.056 & -0.134 & 0.75 $\pm$ 0.27 & 0.63 $\pm$ 1.50 & $107 \pm 48$ & $596 \pm 1504$\\
\enddata
\tablecomments{Ages from \cite{dos_Santos:2016}. [Fe/H] $\pm\ 0.005$ from \cite{brewer:2016a}. [C/H] $\pm\ 0.07$ and [O/H] $\pm\ 0.09$ abundances from \cite{Rice_Brewer_2020} and Polanski et al. (in prep.); *Solar age and isotope ratios from \citep{ayres:2013}}
\label{tab:age,ab,results}

\end{deluxetable*}

\subsection{Kobayashi et al. (2011) Model}
This model includes nucleosynthetic yields from Type Ia supernovae \citep{Nomoto_1997}, updated yields from Type II supernovae and hypernovae \citep{Kobayashi_2006}, and AGB stars \citep{Campbell_Lattanzio_2008, Karakas_2010}. As is evident in this particular model, GCE models that neglect the yields of massive, fast-rotating stars confirm a secondary production mechanism for $^{13}$C which results in a very high ($\sim 10^3$) $^{12}$C/$^{13}$C ratio at low metallicity and is much higher than the solar value $\sim 90$ \citep{kobayashi:2011}. Because the chemical enrichment timescale of the galactic halo is longer than in the solar neighborhood, there is a higher $^{12}$C/$^{13}$C ratio predicted in the halo due to a significant contribution from low-mass AGB stars. In Figure ~\ref{fig:GCE Model Comparison} (top left panel), we see that this model accurately predicts the solar $^{12}$C/$^{13}$C ratio, but overestimates it for stars in the solar neighborhood like the solar twins from the \cite{botelho:2020} sample. The $^{12}$C/$^{13}$C ratio measured here for HIP 85042 (101 $\pm$ 17) and HIP 102040 (105 $\pm$ 48) also fit the model fairly well. The other four solar twins (HIP 29432, HIP 42333, HIP 77052, HIP 79672) demonstrate a significantly lower $^{12}$C/$^{13}$C than model predictions as well as the solar twins in the \cite{botelho:2020} sample. Overall, the decreasing trend in $^{12}$C/$^{13}$C over metallicity is consistent with this GCE model.

All six of our solar twin $^{16}$O/$^{18}$O measurements agree with the \cite{kobayashi:2011} model predictions within the uncertainties, even HIP 102040 and HIP 29432 for which we report the lowest S/N measurements. See Figure \ref{fig:GCE Model Comparison} (top right panel).

\subsection{Prantzos et al. (2018) Model}
All the major isotopes of the multi-isotopic elements up to Fe ($^{12}$C, $^{14}$N, $^{16}$O, $^{20}$Ne, $^{28}$Si, $^{32}$S, $^{36}$Ar, $^{40}$Ca, $^{54}$Cr, $^{56}$Fe) are reproduced to better than 15\% and, in most cases, to better than 10\% in this set of GCE models. \cite{prantzos:2018} provides both a baseline model -- which includes yields from low to intermediate mass stars, massive stars, and rotating massive stars -- and a second model which considers yields from non-rotating massive and low-to-intermediate mass stars only. These models use the metallicity-dependent yields from \cite{Cristallo_2015} and rotating and non-rotating stellar yields from \cite{Limongi_Chieffi_2018}. The inclusion of rotating star yields in the baseline model significantly reduces the $^{12}$C/$^{13}$C ratio at low metallicities ($^{12}$C/$^{13}$C $< 1000$ for [Fe/H] $> -2.0$). Even so, the $^{12}$C/$^{13}$C ratios observed in solar twins with near-solar metallicities are significantly lower than the baseline model's predictions. Therefore, in Figure \ref{fig:GCE Model Comparison} (top/bottom left panels) we plot only the secondary model, not the baseline model. The observed solar twin $^{12}$C/$^{13}$C ratios match the secondary model slightly better. It is worth noting that the $^{12}$C/$^{13}$C measurements presented here show the same trend, decreasing over metallicity, as the GCE models, contrary to the \cite{botelho:2020} solar twin sample. We will explore this issue further in Section 6.4 by comparing the carbon isotope ratios in our six solar twins (HIP 29432, HIP 42333, HIP 77052, HIP 79672, and HIP 102040) to the \cite{botelho:2020} $^{12}$C/$^{13}$C measurements of the same stars.

While these models slightly overestimate $^{12}$C/$^{13}$C ratio observed in solar twins, they demonstrate a slight underestimate for $^{16}$O/$^{18}$O ratios. In the top right and bottom right panels of Figure \ref{fig:GCE Model Comparison}, we observe that five out of the six solar twin measurements of the oxygen isotope ratio presented here, including the Sun with $^{16}$O/$^{18}$O $\sim 511$ \citep{ayres:2013}, show ratios greater than the model predictions of $^{16}$O/$^{18}$O $\sim 300$ for near-solar metallicity stars. It is worth noting that this particular GCE model set would still underestimate the $^{16}$O/$^{18}$O ratio even if we eliminate the lowest SNR measurements in our sample.

\subsection{Romano et al. (2017) Model}
The four carbon isotope ratio models incorporate different combinations of stellar yields. Models 1 and 2 use the same nucleosynthetic prescription for low- to intermediate-mass stars (LIMS) and massive stars \citep{Karakas_2010, Nomoto_2013}; however, Model 2 additionally includes super-AGB star yields \citep{Doherty_2014a, Doherty_2014b}. Model 3 keeps LIMS yields from \cite{Karakas_2010} but pulls the massive star nucleosynthesis prescription from multiple sources \citep{Meynet_Maeder_2002, Hirschi_2005, Hirschi_2006, Ekstrom_2008}. Finally, Model 4 adds super-AGB star yields \citep{Doherty_2014a, Doherty_2014b} to the other nucleosynthesis prescriptions from Model 3. In the bottom left panel of Figure \ref{fig:GCE Model Comparison}, notice that Models 1 and 3 best reproduce solar data \citep{ayres:2013}, while inclusion of super-AGB star carbon synthesis (Models 2 and 4) results in an underestimate of the solar $^{12}$C/$^{13}$C ratio. Nonetheless, all four models predict the current $^{12}$C/$^{13}$C ratios in the solar neighborhood, in agreement with local ISM values, within the errors. Any discrepancies between solar abundances and the local ISM values are typically attributed to the Sun's migration to its current position from a birthplace closer to the galactic center.

If we consider only the $^{12}$C/$^{13}$C ratios from our six solar twin sample, we observe that Models 1 and 2, without fast-rotator yields, perform better and predict the carbon isotope ratios of the entire sample within the errors. Although Models 3 and 4 could, potentially be used to describe the isotopic abundance evolution of younger of solar twins (Age $< 6$ Gyr) particularly those in the \cite{botelho:2020} sample, they do not quite agree with the measurements made in our oldest star (HIP 85042, Age $= 6.66$ Gyr) nor in the youngest (HIP 42333, Age $= 1.01$ Gyr).

In terms of the isotopic oxygen ratio (Figure \ref{fig:GCE Model Comparison}, bottom right panel), Models 1 and 2 also reproduce the $^{16}$O/$^{18}$O ratios measured in the Sun and along the Galactic disc. These two models produce the same $^{16}$O/$^{18}$O evolution as they only differ in the treatment of super-AGB stars, and oxygen is mainly produced in massive stars, not in AGB stars. Older stars (Age $> 6$ Gyr) are predicted to be more $^{18}$O- poor than the Sun while younger stars (Age $< 6$ Gyr) are expected to have little $^{18}$O enrichment relative to the Sun. The older stars in the sample, HIP 85042 and HIP 29432, appear to be enriched in $^{18}$O relative to solar values with $^{16}$O/$^{18}$O ratios lower than the youngest star, HIP 42333, which has a near-solar ratio. 

\subsection{Comparing Carbon Isotope Ratios}\label{subsec:botelho_comp}
The slight $^{13}$CO overestimate (recall we measure $1.10 \pm 0.04$ and $1.03 \pm 0.42$ x Solar abundances for $^{13}$CO and C$^{18}$O respectively \text{in a solar spectrum}; see Section ~\ref{subsec:ab_calc_Sun_ex}) mentioned previously led us to compare our solar twin $^{12}$C/$^{13}$C ratios to archival measurements \citep{botelho:2020}. Using this archival sample, we compare our $^{12}$C/$^{13}$C measurements to a sample of stars 10x larger, with isotopic abundance measurements made with optical CH features rather than infrared CO. This allows us to examine $^{12}$C/$^{13}$C trends across different sample sizes and examine the efficiency of this isotopic abundance analysis using optical \citep{botelho:2020} vs. infrared spectra (this paper).

\begin{figure}[h!tp]
    \centering
    \includegraphics[width=0.5\textwidth]{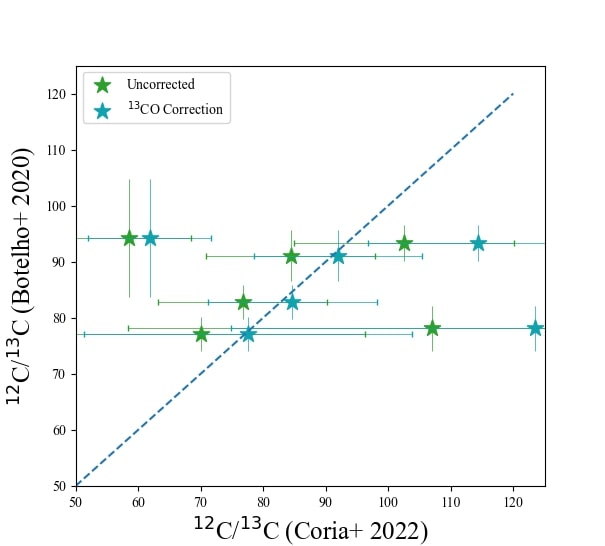}
    \caption{Comparing $^{12}$C/$^{13}$C ratio measurements for our solar twin sample with archival measurements \citep{botelho:2020}. The closer measurements from each study are to each other, the closer they lie to the blue dotted line. All solar twin measurements are consistent within the uncertainties except for HIP 42333 for which we report a significantly lower $^{12}$C/$^{13}$C ratio ($\sim 58$) than in the previous study ($\sim 94$). Refer to Section ~\ref{subsec:botelho_comp} for more details.}
    \label{fig:Archival_Comp}
\end{figure}

In Fig. \ref{fig:Archival_Comp}, all $^{12}$C/$^{13}$C measurements agree within the uncertainties except the one for HIP 42333. Although a mismatch in one out of six stars may not be unexpected, this discrepancy with the \cite{botelho:2020} measurement of the same star is surprising: disagreement may be expected for our lower signal-to-noise measurements such as HIP 29432 with S/N = 16.6, but not for HIP 42333 with a higher signal-to-noise ratio (S/N = 36.2) and a well-behaved line profile. After applying a $^{13}$CO correction of -0.1 x Solar to our overestimated abundances, we recalculate the $^{12}$C/$^{13}$C ratios and plot them in Fig. \ref{fig:Archival_Comp}. The correction brings our four sub-solar $^{12}$C/$^{13}$C ratios closer to archival values. Our two stars with super-solar $^{12}$C/$^{13}$C ratios move further from archival values but still agree within the uncertainties.

The smaller uncertainties for the \cite{botelho:2020} sample are likely due to the higher resolution and S/N of their solar twin spectra. Their sample of solar twins uses HARPS spectra which cover 3780-6910 \AA $\ $ at a resolution R $\sim 115,000$ and reach a S/N of approximately 800 per pixel. These values are significantly higher than what we obtain in the M band with IRTF/iSHELL at a resolution of R $\sim 60,000$: We achieve S/N of just $\sim 20-40$ per pixel. It appears that the $^{12}$C/$^{13}$C ratio may be easier to measure, both in efficiency and precision, using CH spectral features contained in optical spectra. However, there are no ideal oxygen-bearing molecules with spectral features accessible at optical wavelengths. Therefore, measuring the $^{16}$O/$^{18}$O ratio remains possible only using infrared CO features.

In the \cite{botelho:2020} solar twin study, they report that the linear fit of $^{12}$C/$^{13}$C as a function of metallicity shows a positive slope of +56.5 $\pm$ 7.2 dex$^{-1}$. The GCE models discussed above, surprisingly, follow an opposite trend with [Fe/H]. Our solar twin sample demonstrates a weak hint of negative trend for the $^{12}$C/$^{13}$C ratio over metallicity (slope = -157 $\pm$ 82 dex$^{-1}$) that is in accordance with the steady $^{13}$C enrichment predicted by GCE models. However, this is still consistent with the trend of \cite{botelho:2020} at the $3\sigma$ level.

The archival sample also explores trends between the $^{12}$C/$^{13}$C ratio as a function of the isochrone stellar age. They show that $^{12}$C/$^{13}$C ratio is marginally correlated with age (slope of +0.614 $\pm$ 0.250  Gyr$^{-1}$). In our sample, we observe a positive correlation between the $^{12}$C/$^{13}$C ratio and isochrone age with a slope of +6.65 $\pm$ 2.15 Gyr$^{-1}$ that is again consistent with that of \cite{botelho:2020} at $3\sigma$. Both samples are consistent with an overall decrease of the $^{12}$C/$^{13}$C ratio over time.

\section{Conclusions}\label{sec:conclusions}
Our high resolution spectra (R $\sim$ 60,000) from IRTF/iSHELL have made it possible to successfully measure the $^{12}$C/$^{13}$C and $^{16}$O/$^{18}$O ratios in a sample of solar twin stars, the latter of which have never been observed before. 

Our analysis of $^{13}$CO and C$^{18}$O in HIP 102040 and HIP 29432 in particular, the solar twins with the lowest S/N spectra, exemplify the challenge of detecting isotopologues in stellar photospheres. Low S/N spectra result in uncertainties that rival the magnitude of the stellar line profiles themselves. We further observe that the CO isotopologue lines are greatly deformed in these sub-optimal spectra, and this leads to distorted stacked line profiles that do not resemble the Gaussian shape of their stellar model counterparts. These effects are visibly more pronounced in the weaker C$^{18}$O lines (Fig. ~\ref{fig:C_18_O Line Profiles}) and thus make the C$^{18}$O abundance measurements significantly more difficult to make than the $^{13}$CO. 

We also find that both K band and M band spectra are amenable to CO isotopologue analysis depending on the observing time available and the target star's brightness. In their pioneer analysis, \cite{tsuji:2016a} utilized K band spectra to measure $^{12}$C/$^{13}$C in a sample of M dwarf stars. Their study was notably hindered by the low spectral resolution of their K band spectra which resulted in significant blending of the inherently weak absorption lines from the CO overtone band. Our analysis instead targeted the much stronger CO absorption lines in the fundamental rovibrational band. M band spectra are ideal because they provide access to the CO fundamental rovibrational band that produces the strongest minor isotopologue signatures. This is more practical for brighter targets, however, because M band observations require much more observing time to attain the necessary S/N than K band observations. K band observations may be the only practical choice for fainter stars, but the weaker lines in CO overtone band make minor isotopologue detection more difficult.

Nonetheless, the solar twin isotopic ratio measurements fit relatively well compared to the GCE models, within the uncertainties. Since GCE models are generally tailored to solar abundances, we do not expect much deviation from GCE model predictions, at least not from a population of solar twin stars. This isotopic abundance analysis in solar twin stars is a ``pilot study" for similar analyses in GKM-type dwarf stars. The truly interesting isotopic abundance science lies in low metallicity stars.

\subsection{Exploring Isotope Ratios in the Low-Metallicity Regime}
While there has been some progress in measuring the $^{12}$C/$^{13}$C ratio in near-solar metallicity dwarf stars, the more interesting results come from low-metallicity stars. Low metallicity stars provide us a glimpse into the chemical composition of the early galaxy, and thus provide key observational constraints to GCE models. Unfortunately, there are virtually no $^{12}$C/$^{13}$C ratio measurements in the literature for these metal-poor dwarf stars. However, a recent measurement by \cite{spite:2021} shows a sub-solar $^{12}$C/$^{13}$C ratio (27 $<$ $^{12}$C/$^{13}$C $<$ 45) in metal-poor ([Fe/H] = -2.59) dwarf star HD 140283 (log g = 3.70) suggests a much higher $^{13}$C abundance in the early galaxy than is predicted by GCE models. This one carbon measurement, however, makes it difficult to confirm the CNO isotopic composition of the early galaxy. Measurements of the $^{18}$O abundance are significantly more difficult to make in these metal-poor dwarf stars; however, a sufficiently high resolution and high S/N spectrum may reveal the elusive C$^{18}$O features. Regardless of the target or method, we need a larger database of $^{12}$C/$^{13}$C and $^{16}$O/$^{18}$O ratios covering a wide range of stellar metallicities to unveil the mysteries of stellar mixing processes, stellar nucleosynthesis, and the isotopic composition of the early galaxy.

Furthermore, inspection of the GCE models in Section 6 demonstrates that the predicted evolution of major isotope abundances fits solar values quite well; however, evolution of the minor isotopes does not fit solar abundances nearly as well. Because minor isotopes are more fragile than their primary isotopes and destroyed by typical stellar processes, it is difficult to predict minor isotope evolution over time through GCE models. The only hope, then, is to determine the correct [Fe/H] dependence of these minor isotopes. To better explore the production of secondary carbon and oxygen isotopes, we need to expand the isotopic abundance database beyond solar twin stars and measure these isotope ratios in a sample of low-metallicity stars.

\subsection{Isotope Ratios in the Low-Mass Regime}
In addition to being good GCE model calibrators, isotopic carbon and oxygen isotopes may also prove to be good tracers of planetary formation, migration, and atmospheric evolution. Of the 5,044 exoplanets discovered to date (NASA Exoplanet Archive), only one has a $^{12}$C/$^{13}$C ratio measurement: TYC-8998-760-1 b, a young accreting super-Jupiter \citep{zhang_2021a}. While exciting, this new sub-solar carbon isotope ratio does not have a complementary host star measurement which makes it difficult to definitively identify a link between host star abundances, planetary abundances, formation location, and migration \citep{reggiani:2022}. It is also believed that the different formation mechanisms between brown dwarfs and super-Jupiters (e.g. gravitational collapse vs. core accretion) may produce distinct $^{12}$C/$^{13}$C signatures in their atmospheres \citep{zhang:2021b}. Thus, in addition to stellar and planetary isotopic abundances, brown dwarf isotopic abundances introduce yet another important piece of evidence in the planet formation puzzle. 

The first exoplanetary and brown dwarf $^{12}$C/$^{13}$C measurements \citep{zhang_2021a, zhang:2021b}, are likely to be the first of many. Beyond these studies of such giant, H-dominated bodies, CO and H$_2$O isotopologue bands may be detectable even in terrestrial exoplanet atmospheres (orbiting late-type M dwarfs like TRAPPIST-1) using JWST transit transmission spectra throughout the near infrared (1-8 µm), especially at 3-4 µm \citep{lincowski:2019}. JWST and next-generation ELTs may therefore be capable of detecting isotopologues in exoplanet atmospheres ranging from super-Jupiters down to terrestrial-size planets \citep{lincowski:2019}. For now, however, planetary CNO isotope measurements should focus on super-Jupiters and their host stars. Short-period super-Jupiters close to their host stars will provide isotopic abundances within the CO snowline using transmission spectroscopy while bright super-Jupiters, like TYC-8998-760-1 b, provide isotopic abundances outside the CO snowline using a combination of spectroscopy and direct imaging techniques.

\subsection{Future Prospects for Other Isotopes}
Other isotopic abundance ratios for nitrogen, magnesium, silicon, and titanium have been measured previously in giant stars, but because internal processes change these ratios throughout a giant star's lifetime, they do not provide the same constraints on GCE or planetary formation mechanisms as dwarf stars do. Thus, the next important step in building an isotopic abundance database is to go beyond CNO isotopes and measure those previously studied in giant stars and those with the greatest implications for exoplanet formation and composition. Elemental nitrogen abundances are routinely measured in dwarf stars, but the $^{14}$N/$^{15}$N ratio remains unstudied. Nitrogen isotopes would complete the CNO isotopic abundance trifecta, but there are practically no nitrogen isotope measurements for dwarf stars in the literature despite the extensive research done by GCE modelers to predict the evolution of the $^{14}$N/$^{15}$N \citep{Romano:2017}. Optical $^{12}$C$^{15}$N absorption features are sensitive to the $^{15}$N abundance in giant stars \citep{hedrosa:2013} and may be useful in measuring the dwarf star $^{14}$N/$^{15}$N ratios.

Past measurements of the ($^{25}$Mg,$^{26}$Mg)/$^{24}$Mg ratios in cool dwarf stars show that there is a decreasing trend over [Fe/H] which is in accordance with GCE models \citep{Yong:2003}. Further analysis of cool thick-disk and halo stars shows a different trend over metallicity that requires increased $^{25}$Mg and $^{26}$Mg production. This may be attributed to intermediate mass asymptotic giant branch stars \citep{Yong:2003}, and so GCE models can still gain important constraints from Mg isotopes. Silicon abundance measurements in a sample of evolved M-type stars also make good probes of the evolution of metallicity in the ISM. There is a tentative correlation between the $^{29}$Si/$^{30}$Si and the mass-loss rates of these evolved stars which acts as a proxy of stellar age \citep{Peng:2013}. Mg and Si isotopes in exoplanet host stars also may provide useful constraints for exoplanet composition \citep{suarez_andres:2018} and so they provide another key set of isotopes to target for the stellar and planetary isotopic abundance database.

Titanium is also a good target for isotopic abundance studies with features that can be measured in GKM dwarfs stars and potentially, exoplanet atmospheres as well. Titanium has five stable isotopes ($^{46-50}$Ti) with about $25\%$ of its abundance partitioned among the minor isotopes--much greater than the relative abundances of the minor H, C, and O isotopes (all $\sim 2\%$) \citep{Serindag:2021}. This makes Ti isotopes more accessible for challenging observations due to their higher relative abundances compared to other elements. Oxygen and silicon burning in massive stars is responsible for the production of $^{46}$Ti and $^{47}$Ti while $^{48}$Ti, $^{49}$Ti, and $^{50}$Ti are produced mainly in Type Ia and Type II supernovae \citep{Hughes:2008}. This isotope fractionation due to different production mechanisms makes dwarf star Ti isotopes great candidates for testing GCE models. There have been previous measurements of Ti isotope ratios in M dwarf stars of near-solar metallicity with good GCE agreement \citep{Chavez:2009}, but as we have expressed in this paper, the more interesting science lies in sub- and super-solar metallicity stars. In terms of exoplanets, simulations of Ti isotope measurements in exoplanet atmospheres show that an hour of observing time on 8-meter-class telescopes is sufficient to reveal Ti isotopes in the atmospheres of wide-separation super-Jupiters \citep{Serindag:2021}.

\subsection{Concluding Remarks}
In conclusion, this isotopic abundance analysis in solar twin stars served as a ``pilot study" for similar analyses in GKM-type dwarf stars of various ages, metallicities, and even for exoplanet host stars. The agreement between the $^{12}$C/$^{13}$C and $^{16}$O/$^{18}$O measurements made in this paper and GCE model predictions demonstrate the accuracy and precision of calculating isotopic abundances in solar twin stars. The next important step is to fine-tune this process for M dwarf and low-metallicity stars because these stars have the most to benefit from new abundance-age relationships. Furthermore, as the predominant type of exoplanet host stars, M dwarfs provide the best testing sites for abundance-formation relations and planetary isotopic abundance measurements. Together, these stellar and planetary isotopic abundances may eventually unveil ``the missing links" between exoplanet formation mechanisms and exoplanet atmospheric evolution, chemical clocks and stellar ages, and the chemical evolution of the galaxy. 

\clearpage
\bibliography{STCO}{}
\bibliographystyle{aasjournal}



\end{document}